# Transmission of Waveforms Determined by 7 Eigenvalues with PSK-Modulated Spectral Amplitudes

Henning Bülow, Vahid Aref, Wilfried Idler

Nokia Bell Labs, Lorenzstr. 10, 70435, Stuttgart, Germany henning.buelow@nokia.com

**Abstract**  2-ns waveforms with 7 eigenvalues and their QPSK-modulated spectral amplitudes were optimized by taking constraints of link, transmitter, and receiver into account. In experiment these signals were transmitted with a BER of 3.2E-3 over 1440-km of NZ-DSF fiber spans.

**Introduction**

In recent years the modulation of the nonlinear spectrum or detection in this nonlinear domain has been re-discovered and refined for transmission over the fiber in the nonlinear regime. Such schemes are called nonlinear frequency division multiplexing (NFDM)[1,2]. Following the argumentation of Wahls[3], nonlinear Fourier transform (NFT) based modulation and/or detection is motivated by (1) replacing multi step digital back-propagation (DBP) by a less complex one step NFT scheme, by (2) exploiting the quasi linear propagation behaviour of the nonlinear spectrum for operation beyond the nonlinear threshold, or by (3) using detection schemes working on the nonlinear spectrum rather than on the time domain signal. So far, only limited work has been done on lab implementation of these schemes. Recently, the discrete part of the nonlinear spectrum was modulated by on/off keying of up to 4 eigenvalues[4] or by independently QPSK modulated spectral amplitudes of a 2 eigenvalue soliton[5]. Modulation in the continuous part of the nonlinear spectrum was also demonstrated[6].

In this paper, we present the design of pulses with a high count of 7 eigenvalues and QPSK modulation of spectral amplitudes. The eigenvalues are not all located on the imaginary axis as it was the case in the papers before. For the first time, we demonstrate generation, transmission, and detection of such 7-eigenvalue signals in experiment.

**NFDM Signal Design**

Our goal was to design a set of finite eigenvalues (carriers), $\lambda_i$, where the information bits are modulated over spectral amplitudes, $q_d(\lambda_i)$ independently, provided that the following conditions are fulfilled:

(i) Distance between eigenvalues is large enough for reliable detection in Rx in the presence of ASE noise, and numerical limitation of current NFT algorithms. (ii) The largest pulse-width (duration) of all pulses becomes minimum (highest transmission rate) (iii) The largest bandwidth (BW) of all pulses becomes minimum in Tx and Rx (highest spectral efficiency) (iv) No inter-symbol interference (ISI) between adjacent pulses during the transmission (v) Pick-to-average power ratio (PAPR) stays in a practical range.

It is not yet fully clear how to satisfy all these constraints in general, while these conditions are strongly coupled. We relax the condition (iii) to a given BW limit. We limit the BW in Tx to 12.5 GHz, and BW in Rx to 20 GHz assuming the fiber link length is less than 2000 km, including some BW margin for more precise waveform generation and detection. We define BW (pulse-width, respectively) as the frequency (time) interval holding 99% of pulse energy.

We choose a set of 7 eigenvalues as shown in Table 1. The spectral amplitude of each eigenvalue, $\lambda_i$, has QPSK constellation, $q_d(\lambda_i) = |q_d(\lambda_i)| \exp\left(j\frac{\pi}{2}k\right)$. Thus, we modulate at most 14 information bits per pulse.

Distance between eigenvalues is chosen to fulfil conditions (i), (ii) and (iii); on one hand, large difference in imaginary parts may lead to a large pulse-width and large difference in real parts leads to a large BW. On the other hand, small distance between two eigenvalues may also increase both pulse-width and BW.

Since the eigenvalues have different real parts, the pulse will be broadened or contracted during the propagation. To avoid ISI (condition (iv)), we set $|q_d(\lambda_i)| = A_d(\lambda_i)| \exp(2j\lambda_i\tau_0 \times \mathrm{real}\{\lambda_i\})|$. By choosing a positive $\tau_0$, the pulses will first be contracted and then eventually broadened to the same pulse-width and beyond. The link length of reaching to the same pulse-width, $L_{\max}$, depends on $\tau_0$. We optimize $A_d(\lambda_i)$ and $\tau_0$ to have the pulse-width $14\pi$ (for standard nonlinear Schrödinger equation) which scales down to 2 ns in our transmission setup (7 Gbit/s). Using NZ-DSF fiber with $\beta_2 \approx -5.75$ ps$^2$/km and $\gamma \approx 1.6$ W$^{-1}$/km, the optimized $\tau_0$ allows us to have a link about $L_{max} \approx 5600$ km without ISI.

**Table 1:** The discrete spectrum of waveforms $(\lambda_i, q_d(\lambda_i))$ at Tx. $|q_d(\lambda_i)|$ is fixed and $\angle q_d(\lambda_i)$ is QPSK modulated.

| $\lambda_i$ | $0.45j - 0.6$ | $0.3j - 0.4$ | $0.45j - 0.2$ | $0.3j$ | $0.45j + 0.2$ | $0.3j + 0.4$ | $0.45j + 0.6$ |
|---|---|---|---|---|---|---|---|
| $\ln(|q_d(\lambda_i)|)$ | 11.85 | 7.06 | 7.69 | 3.81 | 1.93 | $-0.62$ | $-5.43$ |

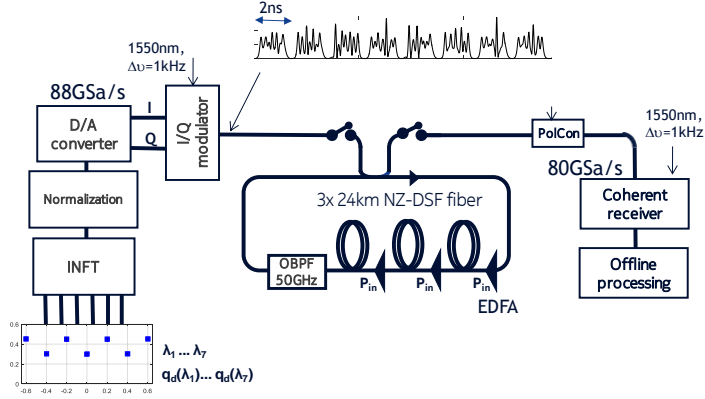
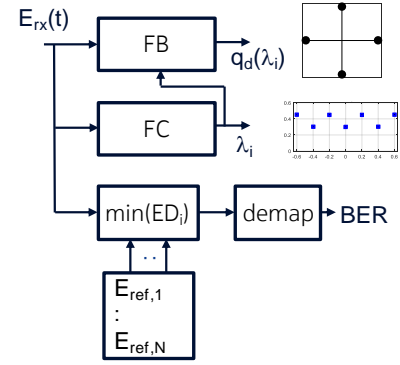

**Fig. 1:** Lab transmission set-up with fiber loop consisting of 3x 24-km NZ-DSF spans and EDFA amplifiers. Transmitter: 88-GSa/s DAC, Receiver: single polarization optical hybrid and 33-GHz bandwidth oscilloscope.

**Fig. 2:** Offline processing of received 2-ns signal traces $E_{rx}(t)$. FB: $q_d$ by a forward-backward NFT algorithm. FC: $\lambda_i$ by Fourier collocation method. min(ED): Decision on min. Euclidian distance.

The optimized $|q_d(\lambda_i)|$ are given in Table 1. Using split-step Fourier method, we simulate the transmission of designed pulses (the random 256 symbols used in experiment) over the NZ-DSF fiber model with the power loss $\alpha = 0.2$ dB/km and span length of 24 km. We numerically observe that we can detect the spectrum error-free (hard-decision over phases) in the range of interest, i.e. 2000 km.

**Experimental set-up and signal processing**

As illustrated in Fig. 1, for experimental evaluation we randomly selected 256 different 2-ns waveforms offline, uploaded these 512-ns sequence into an 88-GSa/s digital-to-analog converter (DAC) driving a Mach-Zehnder IQ modulator. This signal was launched into a NZ-DSF fiber loop set-up with 24-km span length and lumped amplification by EDFA. With this short fiber length, we have only a small penalty because of the mismatch between lossy multi-span fiber in experiment and lossless fiber model of INFT (modified Darboux transform) and NFT. In the transmission experiments we measured optimum launch powers $P_{in}$ of approx. -2.2 dBm which confirms the theoretical path-averaged value of -1.53 dBm. The transmitted signal is coherently detected and a trace of about 2800 2ns-symbols is sampled by a real-time oscilloscope with 80 GSa/s and 33 GHz bandwidth. Best results were obtained if the polarization was adjusted in the optical domain by a manually adjusted polarization controller (PolCon in Fig. 1).

Fig. 2 shows the two different detection schemes were applied separately for signal detection by offline processing:

(1) On one hand the received 2-ns traces $E_{rx}(t)$ were demultiplexed in the nonlinear spectral domain by determining the eigenvalues $\lambda_i$ (FC block in Fig. 2) first and by calculating the spectral amplitude $q_d(\lambda_i)$ for each $\lambda_i$ (i=1...7) using the forward-backward algorithm[7] (FB block). It exhibits an improved numerical stability and precision compared to the often used Ablowitz Ladik integration[2]. At FB output, we disregarded the magnitude of $q_d$ and determined its phase $\angle q_d(\lambda_i)$ which is expected to be a QPSK phase. For calculation (FC block) of the eigenvalues $\lambda_i$, the Fourier collocation (FC) method[2] was applied. A scatterplot of $\lambda_1 ... \lambda_7$ of 2800 symbols in the complex plain is shown in Fig. 3 (c). We observe a clear distinction of the 7 eigenvalues even though some of them are already perturbed rather considerably at the transmitter.

(2) A further decision process (min(ED)) in the time domain was performed separately based on Euclidian distance. For each possible waveform i, (i=1...2$^{14}$), its discrete spectrum is linearly transformed to Rx. The referece waveforms $E_{ref,i}$ are obtained by taking INFT from transformed spectrum. For each received pulse $E_{rx}(t)$, the reference $E_{ref,i}$ was selected which has the least Euclidian distances $ED_i = \int |E_{rx}-E_{ref,i}|^2 dt$. Even though this min(ED) decision is sub-optimum since noise statistics is expected to be signal dependent, it can indicate on whether the waveforms $E_{rx}$ contain sufficient information for reliable decision on the symbols. BER after symbol to bit demapping are provided as output for min(ED) decision (see Fig. 2).

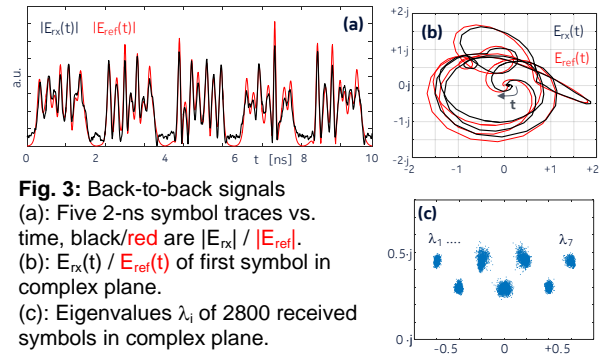

**Fig. 3:** Back-to-back signals
(a): Five 2-ns symbol traces vs. time, black/red are $|E_{rx}|$ / $|E_{ref}|$.
(b): $E_{rx}(t)$ / $E_{ref}(t)$ of first symbol in complex plane.
(c): Eigenvalues $\lambda_i$ of 2800 received symbols in complex plane.

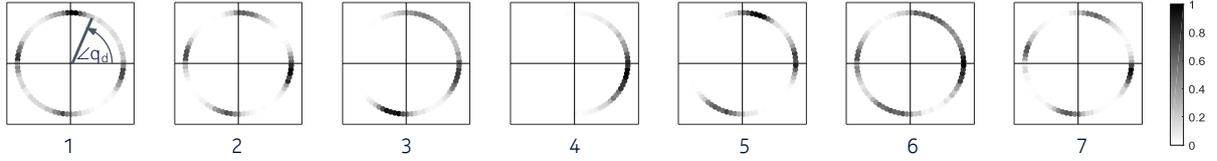

**Fig. 4:** Histograms (2800 symbols) of spectral amplitude phases $\angle q_d(\lambda_i)$ of eigenvalues $\lambda_i$ (i=1…7) measured at transmitter. Gray scale denotes relative occurrence.

**Experimental results**
The magnitude of the first five measured traces $E_{rx}(t)$ of the signal at modulator output are shown in Fig. 3 (a) as black lines. They are compared to the ideal (calculated) signal $E_{ref,i}$ (i=1...5) plotted as red lines. We observe slight deviations at sharp peaks and at zeros between the symbols. Slight but not paramount deviations can also be observed when plotting the first symbol trace $E_{rx}$ in the complex plane, as shown in Fig. 3 (b). We attribute these field perturbations mainly to imperfections of the transmitter[5,8], i.e. limited bandwidth of DAC and modulator drive amplifier, and degraded DAC resolution at high frequencies (ENOB≈5.5bit). Nevertheless the eigenvalues appear in 7 distinct clusters as shown in Fig. 3 (c). The histogram of the measured phases $\angle q_d(\lambda_i)$ at transmitter output are shown as differently shaded gray points on a circle in Fig. 4. The subsets of the 4 QPSK phase states used for signal design are clearly visible. However, the non-vanishing gray points between 90° separated phases (see e.g. histograms 1, 3 and 6 in Fig. 4) already indicate large variance of the phases which we attribute to the lab set-up imperfections and which leads to a high error-floor.

Nevertheless, as summarized in Tab. 2, with min(ED) detection of 2800 waveforms we obtained no error after up to 864-km transmission (12x72km) and BER = $3.2 \cdot 10^{-3}$ after 1440-km (20x72km). We also observed that the (best) launch power $P_{in}$ (bottom row) has the tendency to increase with increasing link length.

**Table 2:** BER and optimum launch power for min(ED) detection vs. link length.

|  | 864 km | 1152 km | 1440 km | 1800 km |
|---|---|---|---|---|
| BER | 0 | $9.1 \cdot 10^{-4}$ | $3.2 \cdot 10^{-3}$ | $1 \cdot 10^{-2}$ |
| $P_{in}$ | -2.3dBm | -2.1dBm | -2.1dBm | -1.8dBm |

**Conclusions**
For the design of waveforms defined by many eigenvalues, several conditions have to be met, such as separation of eigenvalues, minimum pulse width, maximum link length, limited bandwidth at transmitter and receiver, avoiding inter-symbol interference. Under realistic constraints for the lab set-up, we successfully generated 2-ns long waveforms of 7 eigenvalue solitons whose discrete spectra were modulated in phase (PSK).

We transmitted them over up to 1440-km NZ-DSF fiber in lab. A low BER of $3.2 \cdot 10^{-3}$ was obtained by searching for the minimum Euclidian distance of the time domain signal. Even though the low BER confirmed that sufficient information for low BER detection is present in the received symbol waveform, low error-rate decision was not possible when demultiplexing the signal by nonlinear Fourier transform (NFT) and deciding on the phases of the 7 discrete spectral amplitudes $q_d(\lambda_i)$ separately. These observations confirm that 7 eigenvalue solitons were successfully generated and transmitted, but it also shows that the simple concept of demultiplexing eigenvalues by NFT processing in its current form remains challenging for PSK modulated eigenvalues.

**References**
[1] J.E. Prilepsky et al., "Nonlinear inverse synthesis and eigenvalue division multiplexing in optical fiber channels," Physical review letters, 113, 013901, (2014).
[2] M.I. Yousefi et al., "Information Transmission using the Nonlinear Fourier Transform, Part I-III", IEEE Trans. Inf. Theory, vol. 60, no. 7, pp. 4329–4369, (2014).
[3] S. Wahls, "Fiber-Optic Communication using Fast Nonlinear Fourier Transforms," OFC, W3A.1 (2016).
[4] Z. Dong et al., "Nonlinear Frequency Division Multiplexed Transmissions based on NFT," Photon. Technol. Lett., Vol. 27, No. 15, (2015).
[5] V. Aref et al., "Experimental Demonstration of Nonlinear Frequency Division Multiplexed Transmission," Proc. ECOC, Tu.1.1.2, (2015)
[6] Son T. Le et al., "First Experimental Demonstration of Nonlinear Inverse Synthesis Transmission over Transoceanic Distances," OFC, Tu2A.1 (2016).
[7] V. Aref, "Control and Detection of Discrete Spectral Amplitudes in Nonlinear Fourier Transform," Arxiv.org preprint, (2016).
[8] H. Buelow et al., "Experimental Nonlinear Frequency Domain Equalization of QPSK Modulated 2-Eigenvalue Soliton," OFC, Tu2A.3 (2016).